\documentclass[%
 reprint,
 superscriptaddress,
nofootinbib,
 amsmath,amssymb,
 aps,
 prl,
 floatfix
]{revtex4-1}

\usepackage{graphicx}
\usepackage{dcolumn}
\usepackage{bm}
\usepackage{amssymb}
\usepackage{color}
\usepackage[caption=false]{subfig}

\begin{document}

\preprint{APS/123-QED}

\title{First all-sky search for continuous gravitational-wave signals from unknown neutron stars in binary systems using Advanced LIGO data}

\author{%
P.~B.~Covas$^{1}$, Alicia M. Sintes$^{1}$  
}\noaffiliation

\affiliation {Universitat de les Illes Balears, IAC3---IEEC, E-07122 Palma de Mallorca, Spain }


\begin{abstract}

Rotating neutron stars can emit continuous gravitational waves, which have not yet been detected. We present a search for continuous gravitational waves from unknown neutron stars in binary systems with orbital period between 15 and 45 days. This is the first time that Advanced LIGO data and the recently developed \textit{BinarySkyHough} pipeline have been used in a search of this kind. No detections are reported, and upper limits on the gravitational wave amplitude are calculated, which improve the previous results by a factor of 17. 

\end{abstract}

\maketitle






Continuous gravitational waves (CWs) are non-transient and nearly monochromatic gravitational waves (GWs). Neutron stars can emit CWs through a variety of mechanisms, such as rotation with elastic or magnetic deformations (which may be sourced from accretion by a companion), unstable r-mode oscillations, or free precession (see \cite{Lasky} or \cite{Rev} for a recent discussion of different emission mechanisms). Close to their core, these stars have density values equal or higher than the nuclear density, which make them valuable objects to study the unknown equation of state. 
Continuous waves also present an opportunity to test deviations from General Relativity, like searching for extra polarizations of the waves \cite{CWPol} or finding differences between the speed of GWs and the speed of light \cite{CWSpeed}. Several searches for CWs, both from neutron stars in isolated and binary systems, have been previously carried out (see \cite{CWReview} for a recent review of CW searches), although none have conclusively detected a CW signal. Nonetheless, interesting upper limits have been produced which already help to constrain some models of neutron star shape \cite{Def}.

All-sky searches look for emission of CWs from unknown neutron stars in our galaxy, and complement the targeted searches which focus on CWs from known pulsars. Since only a small percentage of the estimated neutron star population has been detected as pulsars, carrying out all-sky searches is important because such a search could discover emission by highly asymmetric neutron stars that have not been detected electromagnetically as pulsars. These searches need to calculate the Doppler modulation (produced by Earth’s rotation and orbit around the Sun) for many sky positions, making their computational cost orders of magnitude higher than the cost of a targeted search. For this reason, the most sensitive methods like matched filtering cannot be used and semi-coherent methods that split the full observation time in smaller chunks (which are incoherently combined) are routinely used. Semi-coherent methods do not recover as much signal-to-noise ratio as coherent methods do, but the number of templates that need to be searched over in order to constrain the maximum mismatch between signal and template is greatly reduced, thus highly decreasing the computational cost of the search. A recent comparison between different semi-coherent methods is shown in \cite{SemiComp}.

All-sky searches for neutron stars in binary systems pose an even more difficult problem, since the parameters that describe the orbit around the binary barycenter also need to be included in the search parameters. These searches are valuable, since approximately half of the known pulsars with rotational frequencies above 20 Hz belong to binary systems. Until recently, there was only one mature semi-coherent pipeline which could carry out this type of search, called \textit{TwoSpect} \cite{TwoSpectMethods}. This pipeline has been used once in a search for CW signals using the S6 and VSR2-3 datasets \cite{TwoSpectResults}, reporting no detections. 

Recently, we developed a new pipeline called \textit{BinarySkyHough} (BSH) \cite{BSH}. This pipeline is an extension of the semi-coherent \textit{SkyHough} pipeline \cite{SkyHough}, which has been used in many past all-sky searches. It replaces the search over the spin-down/up parameter of isolated sources for the three binary orbital parameters characterizing different possible circular orbits. As explained in \cite{BSH}, this is computationally achievable due to both the usage of the massive parallelization which GPUs (Graphical Processing Units) provide and the computational advantages employed by \textit{SkyHough}. Initial tests indicate that the BSH pipeline provides roughly two times more stringent upper limits, although these tests have been done over a smaller parameter space. 
In this letter we present the first application to real data of this new pipeline. No detections are reported, but the improved quality of the datasets and the new pipeline allows us to improve the upper limits by a factor of 17.


\textit{Signal model}.--- A neutron star with an asymmetry around its rotation axis emits CWs, which produce a time-dependent strain that can be sensed with interferometric detectors. 
The amplitude of this signal is given by \cite{Fstat}:
\begin{align}
        h_0 = \frac{4\pi^2G}{c^4} \frac{I_{zz} \epsilon f^2}{d},
        \label{eq:h0}
\end{align}
where $d$ is the distance from the detector to the source, $f$ is the gravitational-wave frequency (equal to two times the rotational frequency), $\epsilon$ is the ellipticity or asymmetry of the star, defined by $(I_{xx}-I_{yy})/I_{zz}$, and $I_{zz}$ is the moment of inertia of the star with respect to the principal axis aligned with the rotation axis.

The time-dependence of the gravitational-wave frequency is given by \cite{BSH}:
\begin{align}
    f(t) = f_0 + f_0\frac{\vec{v}(t)\cdot\hat{n}}{c} - f_0 a_p \Omega \cos{[\Omega(t-t_{\text{asc}})]},   \label{eq:frequencyevoFinal}
\end{align}
where $\vec{v} (t)$ is the velocity vector of the detector, $f_0$ is the gravitational-wave frequency defined at some reference time, and $a_p$, $\Omega$ and $t_{asc}$ respectively represent the projected semi-major axis amplitude (in light-seconds), angular frequency of the binary orbit and time of ascending node (the three parameters describing the binary orbit). This is the frequency-time pattern that we search, which depends on six unknown parameters that need to be explicitly searched over: $f_0$, $\alpha$ (right ascension), $\delta$ (declination), $\Omega$, $a_p$ and $t_{\text{asc}}$.

This model assumes a circular binary orbit, but as discussed in \cite{BSH}, our pipeline remains fully sensitive to signals with eccentricity less than $10^{-2}$. The model also assumes that the neutron star does not suffer any glitches during the observing time, and that the effect of spin-wandering (stochastic variations on the rotational frequency due to the accretion process) as estimated in \cite{SpinWand}, if present, can be neglected. Although we don't explicitly search over a spin-down/up parameter, this search is sensitive to sources with spin-down/up up to $(T_c T_{obs})^{-1} = 4.8 \times 10^{-11}$ Hz/s (where $T_c=900$ s is the coherent time and $T_{obs}=23170808$ s is the time span of the datasets), since sources with this value or lower wouldn't change the frequency-time pattern by more than a frequency bin, thus not producing any observable change. All known pulsars in binary systems have spin-down values lower than this quantity \cite{BSH}.


\textit{Search}.--- To perform the main search we use the BSH pipeline \cite{BSH}. The full Advanced LIGO \cite{Detector} O2 dataset \cite{O2Data} (publicly available in \cite{GWOSC}) is used, comprised of data from the H1 (Hanford) and L1 (Louisiana) detectors without segments that contain epochs of extreme contamination (the used segments are listed in \cite{Segments}, where the files with the ``all'' tag are selected). The O2 run started on November 30 2016 and finished on August 25 2017. The H1 detector suffered from jitter noise, and a separate data stream (which we use) that removes this contamination was created in order to improve the amplitude spectral density of the detector (more details are given in \cite{Cleaning}). The H1 and L1 datasets include artificially added signals, called hardware injections, which help to test the performance of the detectors and the sensitivity of the different search algorithms (although no hardware injections with binary orbital modulation are present). The parameters of the hardware injections are given in \cite{O2AllSky}. Furthermore, these datasets contain several lines and combs, described with more detail in \cite{LinesCombs}. These disturbances, usually narrow in frequency, are problematic because they can imitate and/or mask the signals we are looking for, thus lowering the sensitivity of our pipeline. 

The input data, described as a signal plus additive noise $x(t) = h(t)+n(t)$, is converted to the frequency-domain and kept as a collection of ``Short Fourier transforms'' (SFTs). Each of these SFTs has a coherence time $T_c$ of  900 s, in order to constrain the gravitational-wave signal in a single frequency bin and not lose power to neighbouring bins (due to the two orbital modulations which affect the searched signal) \cite{BSH}. From these constraints, 14788 and 14384 SFTs from H1 and L1 are obtained, making a total of $N_{SFTs} = 29172$ which are analyzed together. 

Table \ref{tab:Search} shows the parameter space that has been searched. We split the search in frequency bands of 0.1 Hz, each of these covering all the sky and the full range of binary orbital parameters. The resolution for each of these parameters is given by \cite{BSH}:
\begin{align}
    \delta f_0 = \frac{1}{T_c}, \quad
    \delta \Theta = \frac{c}{v T_c f P_F}, \quad \delta a_p = \frac{\sqrt{6 m}}{\pi T_c f \Omega}, \nonumber \\
    \delta \Omega = \frac{\sqrt{72 m}}{\pi T_c f a_p \Omega T_{obs}}, \quad 
    \delta t_{\text{asc}} = \frac{\sqrt{6 m}}{\pi T_c f a_p \Omega^2},
    \label{eq:res}
\end{align}
where $v/c=10^{-4}$, $\Theta$ represents both right ascension and declination, $m$ is a parameter which controls the resolution of the binary parameters and $P_F$ the resolution of the sky position parameters. Different values for $m$ and $P_F$ (shown in table \ref{tab:Resolution}) are selected depending on the frequency, in order to have a manageable Random Memory Access usage and a nearly constant number of templates per 0.1 Hz band across the frequency range. 

For each of these bands the main search returns a list with a percentage of the most significant templates ordered by a detection statistic. Our pipeline is divided in two main stages which use different detection statistics (more details are explained in \cite{BSH}). The top $5\%$ templates in each 0.1 Hz band go to the second stage, and the final toplist only contains $0.1\%$ of the templates passed to the second stage. 
The second stage of the search uses a complementary set of SFTs, which is generated from the initial set by moving the initial time of each SFT by $T_c/2$ and creating a new SFT at each new timestamp (if a contiguous set of data of $T_c$ seconds exists). This procedure slightly increases the sensitivity of the procedure as explained in \cite{Sliding}.

After running the main search, a clustering procedure is applied to the returned toplists. This procedure improves the parameter estimation and allows us to reduce the number of candidates that need to be followed-up. For this search we use a clustering distance threshold of $\sqrt{14}$ (as used in past searches), where the distance is defined as:
\begin{align}
d^2 = &\left(\frac{\Delta f}{\delta f}\right)^2 + \left(\frac{\Delta x}{\delta \theta}\right)^2 + \left(\frac{\Delta y}{\delta \theta}\right)^2 \nonumber \\ + &\left(\frac{\Delta a_p}{\delta a_p}\right)^2 + \left(\frac{\Delta \Omega}{\delta \Omega}\right)^2 + \left(\frac{\Delta t_{\text{asc}}}{\delta t_{\text{asc}}}\right)^2.
\label{eq:postdist}
\end{align}
Quantities in the denominator represent the resolution in each dimension given by equations \eqref{eq:res}, and $x$ and $y$ are the Cartesian ecliptic coordinates projected in the ecliptic plane. Clusters are found by calculating the distance between all templates, and keeping a list with indices of members with distances below the threshold. 
Afterwards, the center of each cluster is found as a weighted (by power significance) sum for each of the six parameters. We keep the 3 most significant clusters per 0.1 Hz band, ordered by the maximum detection statistic value of each cluster, only keeping clusters which have at least 3 members. This produces the list of 6000 outliers from the main search, coming from the 2000 frequency bands. 

\begin{table}[tbp]
\begin{center}
\begin{tabular}{ r c c }
\hline
Parameter & Start & End \\
\hline \hline
Frequency [Hz] & 100 & 300 \\
Right ascension [rad] & 0 & $2\pi$ \\ 
Declination [rad] & $-\pi/2$ & $\pi/2$ \\
Period [day] & 15 & 45 \\ 
$a_p$ [s] & 10 & 40 \\ 
Time of ascension [s] & $t_{mid}$ - $P$/2 & $t_{mid}$ + $P$/2 \\ 
\hline
\end{tabular}
\caption{Ranges of the different searched parameters. Period $P$ is given by $2\pi/\Omega$, and $t_{mid}$ is the mean between the starting and ending times of the datasets.} 
\label{tab:Search}
\end{center}
\end{table}

\begin{table}[tbp]
\begin{center}
\begin{tabular}{ c c c }
\hline
Frequency range & $m$ & $P_F$ \\
\hline \hline
$[100,125)$ & 0.4 & 1 \\
$[125,150)$ & 0.8 & 1 \\
$[150,200)$ & 1.4 & 1 \\
$[200,250)$ & 2.4 & 0.75 \\
$[250,300)$ & 3.4 & 0.75 \\
\hline
\end{tabular}
\caption{Resolution parameters at different frequency ranges.}
\label{tab:Resolution}
\end{center}
\end{table}


The next step consists of applying vetoes to these outliers in order eliminate the ones produced by non-astrophysical sources. The first veto that we apply is the \textit{lines veto}, used in many past searches such as \cite{O2AllSky}. This veto calculates the frequency-time pattern for each outlier and checks if it crosses any frequency where there is a known line or comb, listed in \cite{LinesCombs}. After applying this veto only 4937 outliers remain.


In order to follow-up these outliers, we use the strategy of repeating the search in multiple steps with an increased coherence time (still using a semi-coherent approach) and a reduced range of parameter uncertainty. If the outlier is produced by a real astrophysical signal, the detection statistic will keep increasing, while the same behaviour is not expected for Gaussian noise. 
The multi-detector $\mathcal{F}$-statistic (the frequentist maximum-likelihood statistic), derived in \cite{Fstat} and \cite{MultiFstat}, can be used to perform searches with longer coherence times without losing power to neighbour frequency bins. The computational cost of a gridded $\mathcal{F}$-statistic search over six parameters for such a long dataset would be too high, and for this reason we need to use a method with stochastic placement of templates.

The procedure outlined in \cite{Followup} and \cite{Followup2} consists of using a tempered ensemble walker MCMC algorithm (called \textit{ptemcee} \cite{ptemcee}) to draw samples of the $\mathcal{F}$-statistic and converge to the true signal parameters. To use this procedure, the coherent time and the width in each dimension around the cluster center that we want to follow need to be selected. Wider regions will achieve a higher rate of detected signals, since the centers of the clusters can be located at several bins from their true location, but they will incur in higher computational costs because to reach convergence of the MCMC algorithm the number of steps and/or walkers needs to be increased. The same happens with $T_c$: longer times are able to achieve higher sensitivity, but they require more steps to converge.

The behaviour of the follow-up is characterized by adding simulated signals (called injections) to the datasets. We use 4573 injections at 4 different values of $h_0$ and 10 different non-disturbed frequencies. The $h_0$ values are located near the $95\%$ detection efficiency point, which is derived later. Firstly, we run BSH to obtain the clusters for each injection, and then we follow-up with $T_c=60000$ s (the number of segments is 387) the injections whose cluster's centers are within 5 bins of the true parameters (the injections which count as detected by BSH). All but 9 injections are recovered with a semi-coherent $\mathcal{F}$-statistic value $2\mathcal{F}_{sum}$ of more than 2000. This value is used as a threshold for the follow-up of the outliers, which implies a false dismissal of $9/4573 = 0.1\%$.  

From the 4937 outliers, only 27 have $2\mathcal{F}_{sum}$ values above the threshold, listed in \cite{OutliersTable}, grouped around 8 different frequency regions. Before running the next stage of the follow-up, we inspect them carefully. This reveals that for the outliers at 7 of these frequency regions, the detection statistic in one of the detectors is much higher than in the other one, and that most of it is accumulated during a small portion of the run. These outliers can be safely attributed to disturbances which were present for a short time, thus for this reason they are not present in the lines and combs database (since it is obtained by using a mean amplitude spectral density of the full observing run).

The outliers remaining at the frequency region around 190.6 Hz, present similar $2\mathcal{F}_{sum}$ values in both detectors, but also accumulate their statistic during a short portion of the run. After a closer inspection, these outliers seem to be generated by one of the hardware injections present in the data: the recovered parameters make the template closely resemble the frequency-time pattern of the hardware injection during a few days, and due to their huge $h_0$ values, a high value of the detection statistic is accumulated even for such short fractions of the run. Thus, all search outliers are caused by non-astrophysical sources and are vetoed. No astrophysical signals are confidently detected in this search. 

\begin{figure}[tbp]
\includegraphics[width=1.0\columnwidth]{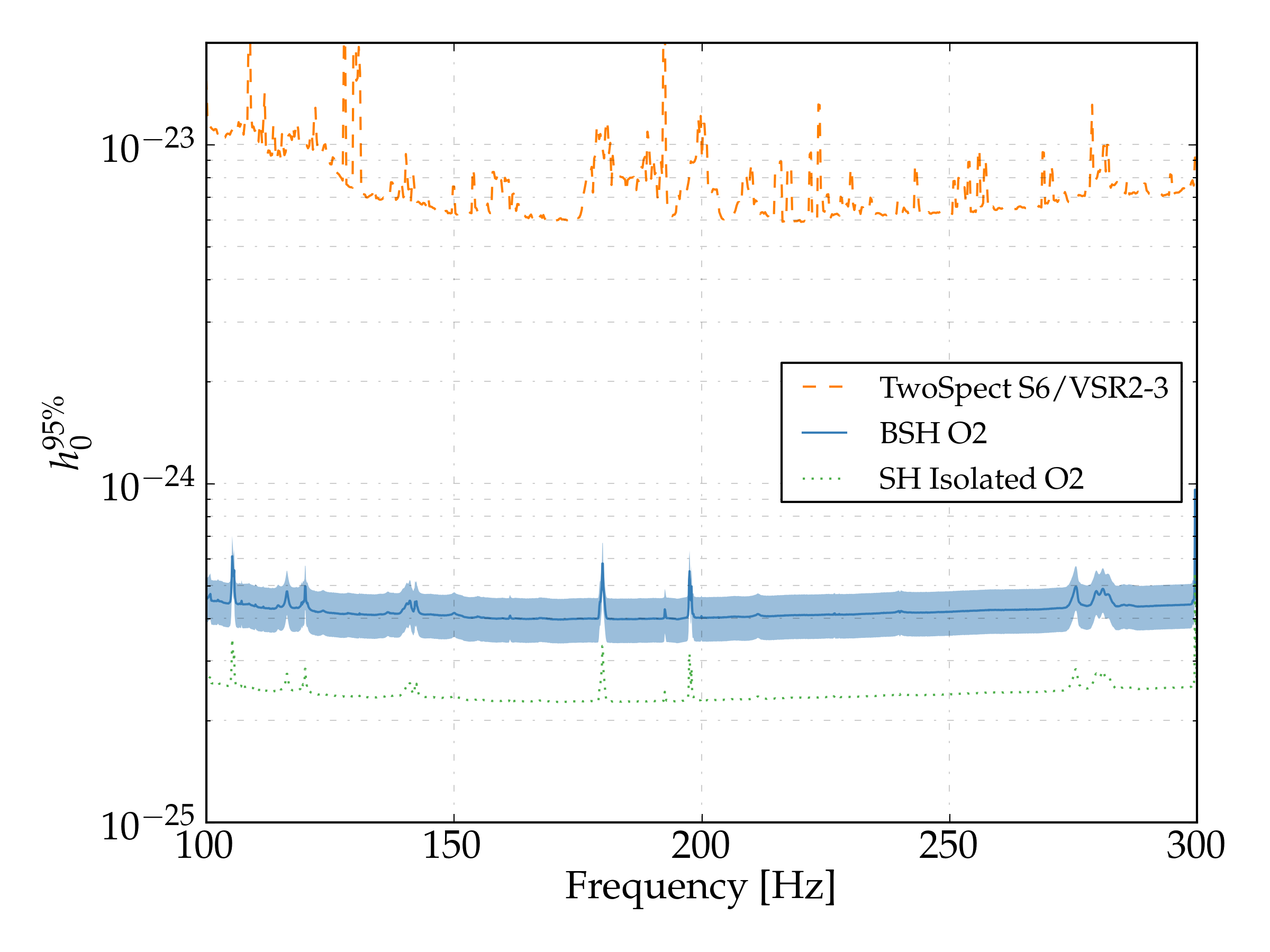}
\caption{Upper limits on the gravitational wave amplitude $h_0$ at $95\%$ confidence to isotropically polarized signals. The middle blue trace with an envelope shows the \textit{BinarySkyHough} results and their uncertainty, the upper dashed-orange trace shows the S6/VSR2-3 results produced by the \textit{TwoSpect} pipeline \cite{TwoSpectResults}, and the bottom dotted-green trace shows the results for the O2 CW all-sky search for isolated neutron stars using the \textit{SkyHough} pipeline \cite{O2AllSky}.}
\label{fig:UL}
\end{figure}

\textit{Results}.--- Although no detections are reported, we set upper limits on the gravitational-wave amplitude. Again, signals are added to the datasets in 10 different non-disturbed frequency bands at 4 different sensitivity depth $\mathcal{D} = \sqrt{S_n}/h_0$ values (where $\sqrt{S_n}$ is the amplitude spectral density), using 300 signals per depth and frequency. For each of them we calculate the efficiency, which is the number of detected signals divided by the number of injected signals. This procedure takes into account the first stage of the follow-up, where only the injections which obtain a $\mathcal{F}$-statistic value above the threshold are counted as detected, and it is also required that the injection cluster has a maximum detection statistic higher than the maximum detection statistic of the third cluster found in that 0.1 Hz frequency band, because otherwise it would not have been detected. Then, at each of the 10 frequency bands a linear fit is done and the $95\%$ sensitivity depth value is found. Although the resolution parameters (as shown in table \ref{tab:Resolution}) decrease with frequency, the sensitivity depth at which we achieve a $95\%$ efficiency is not greatly reduced, as explained in \cite{BSH}. For this reason, we calculate the mean between the 10 frequency bands and use that sensitivity depth to calculate a unique upper limits trace. The result is $\mathcal{D}^{95\%} = 18.5 \pm 2.1$ Hz$^{-1/2}$. 


The upper limits are shown in figure \ref{fig:UL} (they are only strictly valid in frequency bands where lines or non-Gaussianities are not present, a list with the non-valid frequency bands is presented in \cite{OutliersTable}). It can be seen that the lowest gravitational-wave amplitude is located around $4 \times 10^{-25}$ near 170 Hz. This figure shows a comparison with the previous upper limits obtained by analyzing data from S6 and VSR2-3, discussed in \cite{TwoSpectResults}. The sensitivity to $h_0$ scales as \cite{EstSens}:
\begin{align}
        h_0 \propto \frac{\sqrt{S_n}}{T_c} N_{SFTs}^{\sim0.25}.
        \label{eq:h0Sens}
\end{align}
At 150 Hz, for S6 the amplitude spectral density was around $2\times10^{-23}$ Hz$^{-1/2}$, which compared to O2 ($\sim$$7\times10^{-24}$ Hz$^{-1/2}$) gives a factor $\sim$3 of improvement. The S6 run covered a longer calendar period than O2, but the duty cycle was worse so the overall $N_{SFTs}$ factor from each run is comparable. 
Therefore, the improvement of $\sim$17 that can be seen in the figure is due primarily to the improved detector data set as well as using the new BSH pipeline. An important distinction remains that this search covers a much smaller parameter space compared to the previous \textit{TwoSpect} S6/VSR2-3 search. A more complete comparison would need to take this distinction into account. Figure \ref{fig:UL} also shows the previously published results for the O2 all-sky search for CWs from isolated systems using the \textit{SkyHough} pipeline \cite{O2AllSky}. The upper limit results presented here for CWs from sources in binary systems is only a factor of $\sim$2 worse, which is a new achievement for this type of search.


\begin{figure}[tbp]
\includegraphics[width=1.0\columnwidth]{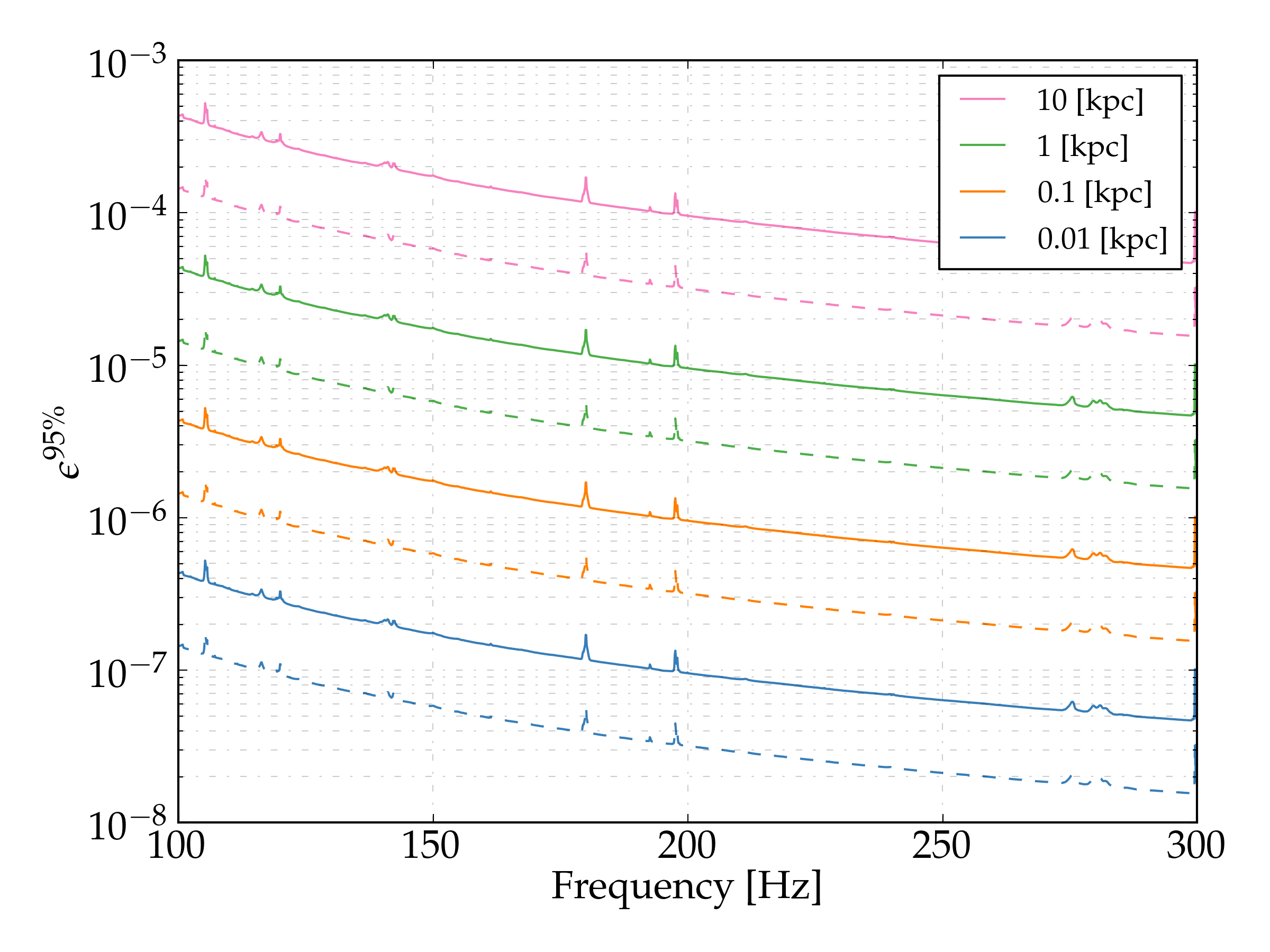}
\caption{Detectable ellipticity at $95\%$ confidence, given by equation \eqref{eq:eps}, as a function of gravitational-wave frequency for neutron stars at 10 pc (bottom trace), 100 pc, 1 kpc and 10 kpc (upper trace) for a canonical moment of inertia $I_{zz}=10^{38}$ kg$\cdot$m$^2$ (regular traces) and $I_{zz}=3\times10^{38}$ kg$\cdot$m$^2$ (dashed traces).}
\label{fig:Reach}
\end{figure}

The $95\%$ upper limits on $h_0$ can be converted to upper limits on ellipticity $\epsilon$ by using equation \eqref{eq:h0}:
\begin{align}
        \epsilon^{95\%} = \frac{c^4}{4\pi^2G} \frac{h_0^{95\%} d}{I_{zz} f^2}.
        \label{eq:eps}
\end{align}
These results are shown in figure \ref{fig:Reach}, where different values for the moment of inertia and distances are used. Assuming the canonical moment of inertia of $I_{zz}=10^{38}$ kg$\cdot$m$^2$, for sources at 1 kpc emitting CWs at 300 Hz the ellipticity can be constrained at $\epsilon< 5 \times 10^{-6}$; at 100 Hz, $\epsilon < 4\times 10^{-5}$, while at 0.1 kpc and 300 Hz, $\epsilon<4\times10^{-7}$. 
If we assume $I_{zz}=3\times 10^{38}$kg$\cdot$m$^2$ (as could be due to higher masses or larger radii), these upper limits are even more stringent, as shown by the dashed traces in this figure. For example, at 0.1 kpc and 200 Hz, $\epsilon<3\times10^{-7}$, while at 0.01 kpc and 300 Hz $\epsilon<2\times10^{-8}$. Several studies indicate that neutron stars should be able to support ellipticities greater than $10^{-5}$ \cite{Def}, making our results interesting in terms of constraining the asymmetry which neutron stars in binary systems have. 


The main search done by the BSH pipeline took 10000 CPU-hours to complete (by using a Power9 8335-GTH + Tesla V100 GPU combination), which is a very small cost. The O2 data could be further searched for signals in other regions of parameter space, both at higher frequencies and at lower and higher orbital periods. This could also be done with the next set of Advanced detectors O3 data, which will have an improved noise floor that will produce even tighter upper limits and enhance the possibilities of detection. 

\vspace{\baselineskip}
The authors want to thank Evan Goetz, David Keitel, Gregory Ashton, and the CW LVC group for multiple discussions and suggestions which improved the quality of this publication. This research has made use of data, software and/or web tools obtained from the Gravitational Wave Open Science Center (https://www.gw-openscience.org) \cite{GWOSC}, a service of LIGO Laboratory, the LIGO Scientific Collaboration and the Virgo Collaboration. LIGO is funded by the U.S. National Science Foundation. Virgo is funded by the French Centre National de Recherche Scientifique (CNRS), the Italian Istituto Nazionale della Fisica Nucleare (INFN) and the Dutch Nikhef, with contributions by Polish and Hungarian institutes.
We acknowledge the support of the Spanish Agencia Estatal de Investigaci{\'o}n and Ministerio de Ciencia, Innovaci{\'o}n y Universidades grants FPA2016-76821-P, FPA2017-90687-REDC, FPA2017-90566-REDC, RED2018-102661-T, RED2018-102573-E, the Vicepresidencia i Conselleria d'Innovaci{\'o}, Recerca i Turisme del Govern de les Illes Balears (grant FPI-CAIB FPI/2134/2018) and the Fons Social Europeu 2014-2020 de les Illes Balears, the European Union FEDER funds, the Generalitat Valenciana (PROMETEO/2019/071), and the EU COST actions CA16104, CA16214, CA17137 and CA18108. The authors are grateful for computational resources provided by the LIGO Laboratory and supported by National Science Foundation Grants PHY-0757058 and PHY-0823459. The authors thankfully acknowledge the computer resources at CTE-POWER and the technical support provided by \textit{Barcelona Supercomputing Center - Centro Nacional de Supercomputaci\'on} (RES-AECT-2019-3-0011). This article has LIGO document number P1900326.

\end{document}


\preprint{APS/123-QED}

\title{Supplemental material for ``First all-sky search for continuous gravitational-wave signals from unknown neutron stars in binary systems using Advanced LIGO data''}

\author{%
P.~B.~Covas$^{1}$, Alicia M. Sintes$^{1}$  
}\noaffiliation

\affiliation {Universitat de les Illes Balears, IAC3---IEEC, E-07122 Palma de Mallorca, Spain }

\maketitle

\section{Follow-up plots}

\begin{figure}[tbp]
\includegraphics[width=1.0\columnwidth]{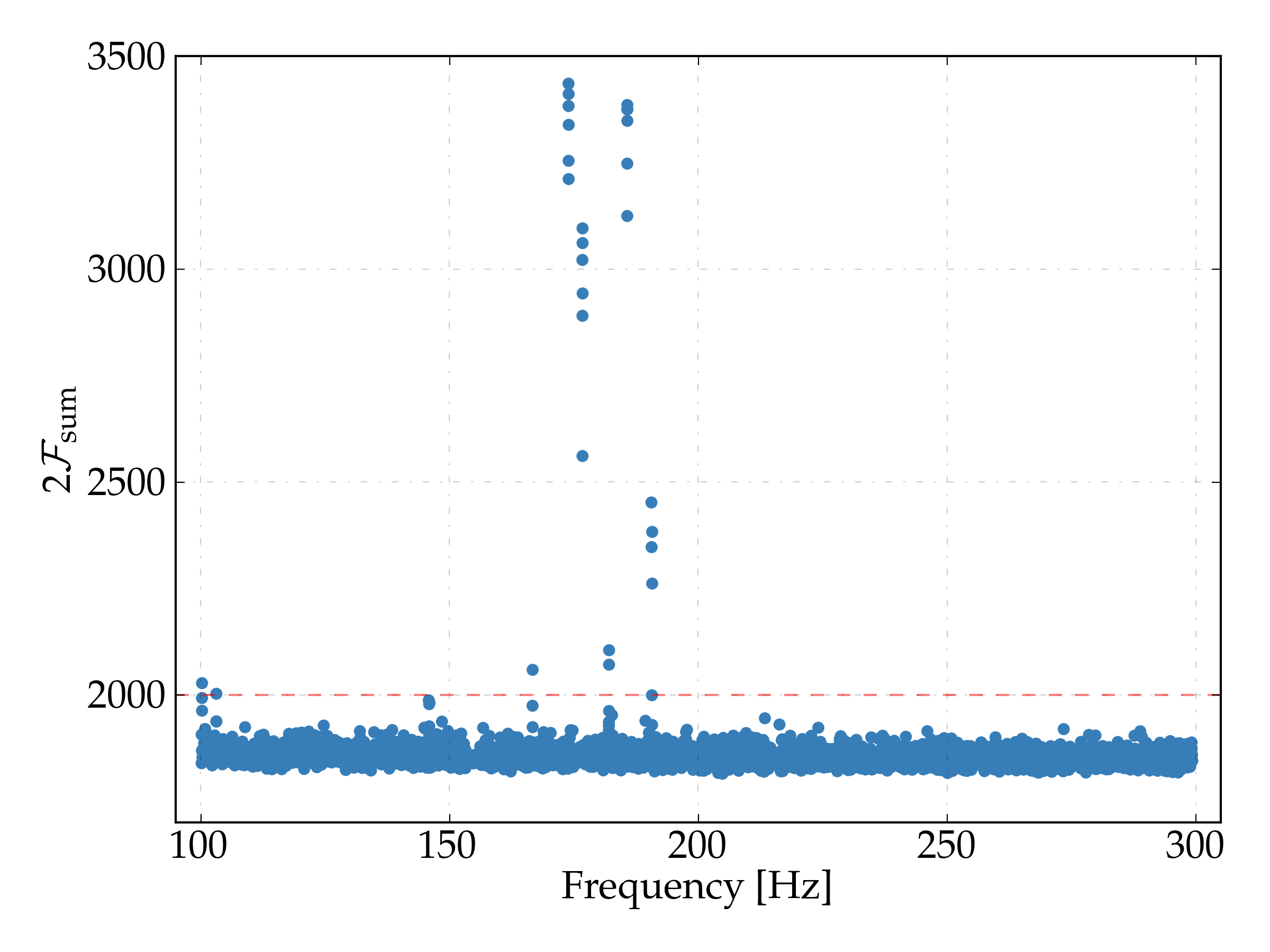}
\caption{$2\mathcal{F}_{sum}$ values for all the 4937 outliers which have been followed-up. The red dashed line marks the threshold at 2000.}
\label{fig:FCl}
\end{figure}

\begin{figure}[tbp]
\includegraphics[width=1.0\columnwidth]{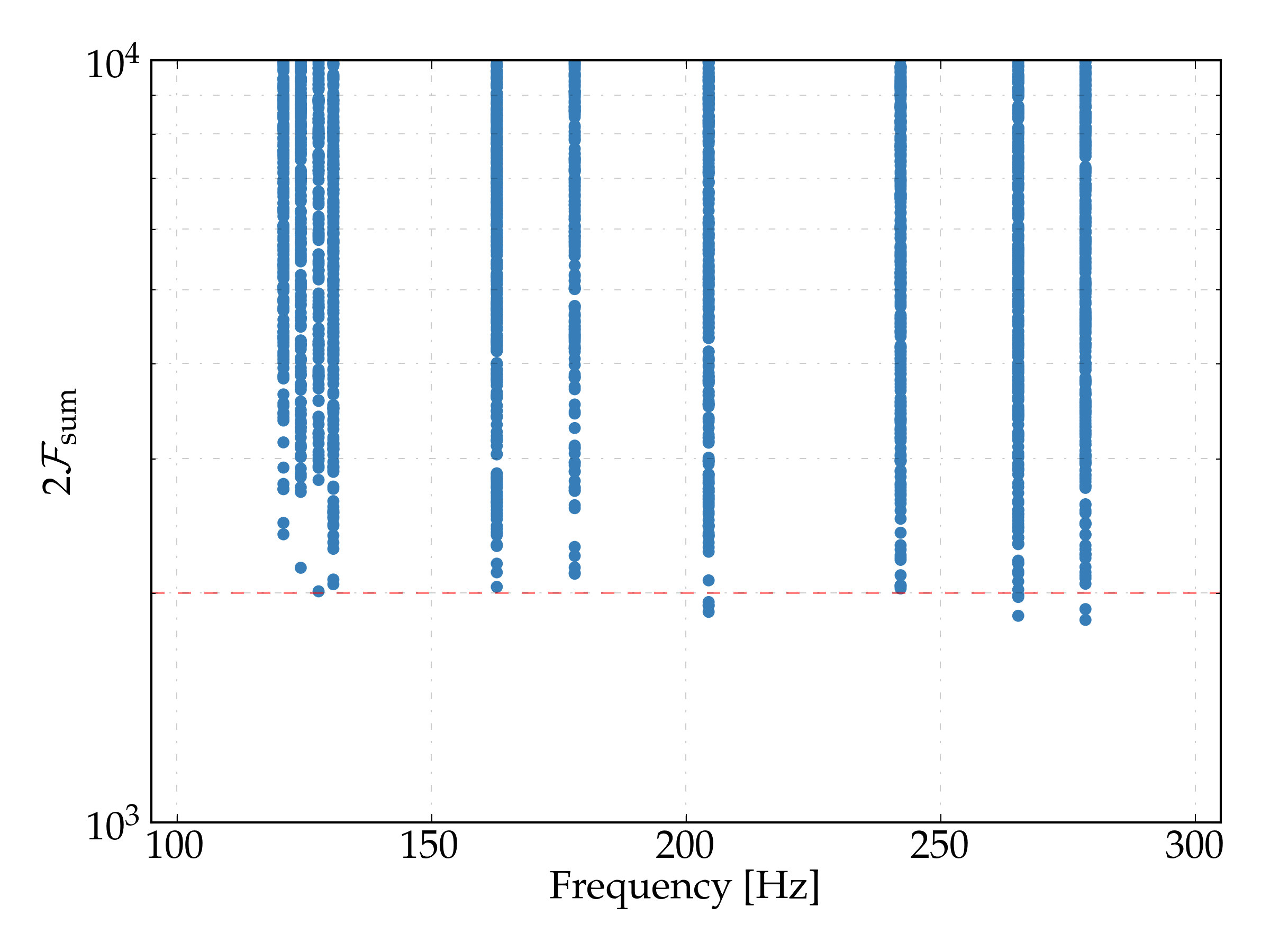}
\caption{$2\mathcal{F}_{sum}$ values for all injections done at 10 different frequency bands, with 4 different sensitivity depth values per frequency band. The total number of injections is 4573, and 9 of them are above 2000.}
\label{fig:FInj}
\end{figure}

We show the $2\mathcal{F}_{sum}$ values for all the outliers in figure \ref{fig:FCl}. Furthermore, figure \ref{fig:FInj} shows the $2\mathcal{F}_{sum}$ values for the injections. The threshold at $2\mathcal{F}_{sum} = 2000$ is marked with a dashed red line in both plots. Due to the small number of outliers which are above 2000, outliers which are slightly below 2000 are also manually followed-up and they also appear in table \ref{tab:SkyHoughOutliers}.

\section{Sensitivity depth estimation}

\begin{figure}[]
\includegraphics[width=1.0\columnwidth]{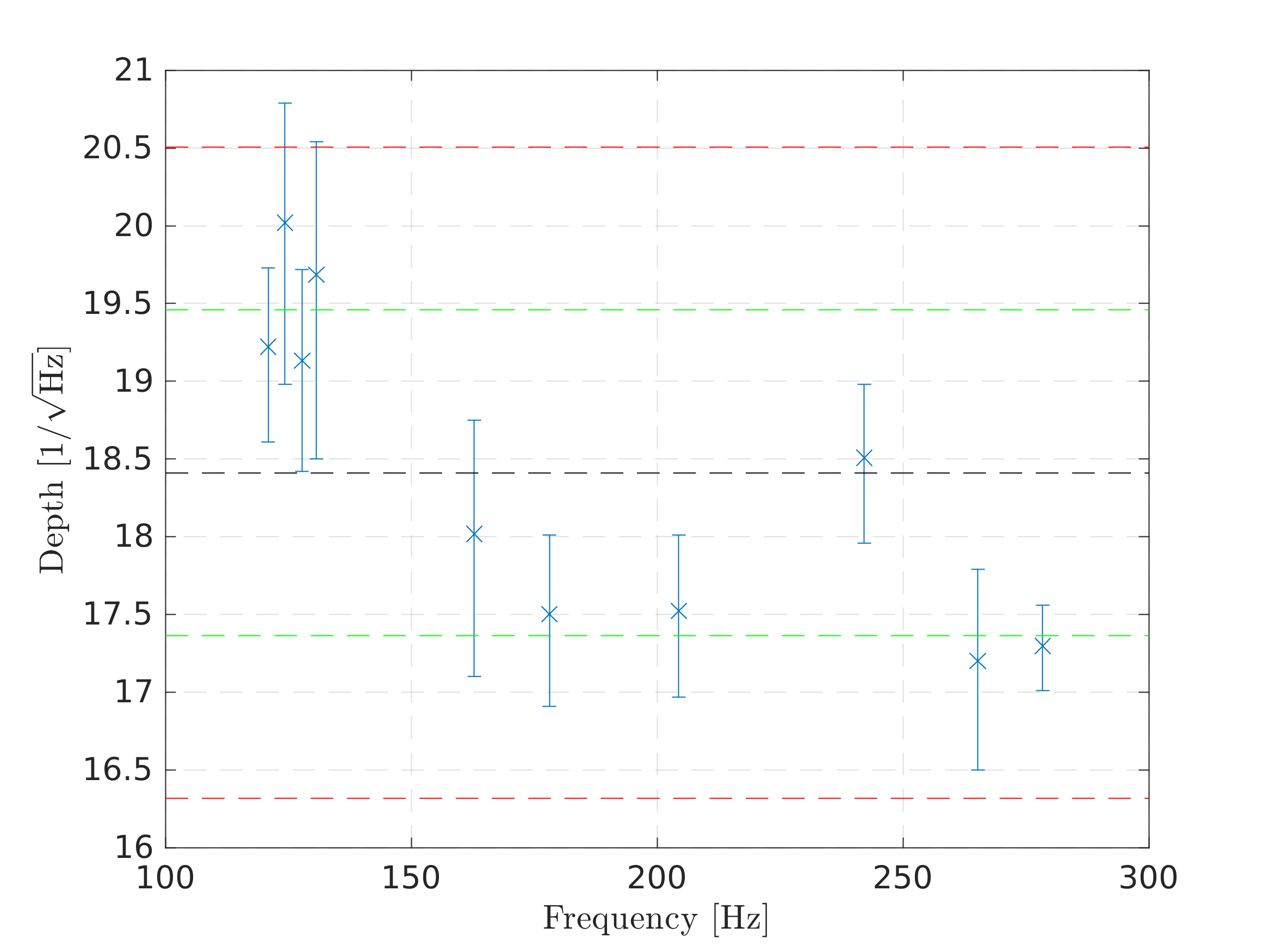}
\caption{$95\%$ sensitivity depth points for the injections done at 10 different frequency bands. The green and red lines mark the position of one and two standard deviations respectively.}
\label{fig:Depth2}
\end{figure}

\begin{figure}[]
\includegraphics[width=1.0\columnwidth]{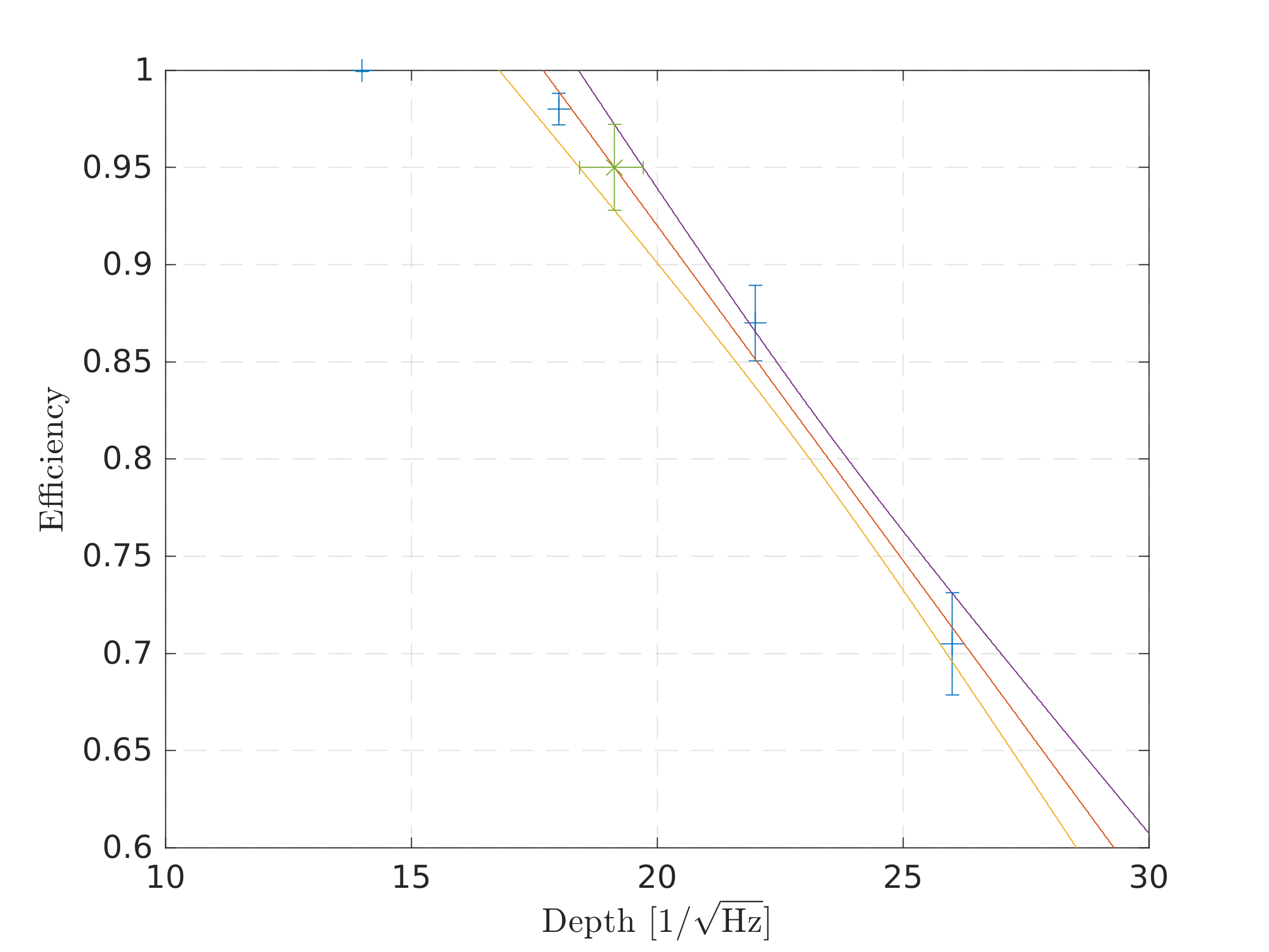}
\caption{Example of a fit at 127.8 Hz. The four blue crosses show the efficiencies obtained at sensitivities of 14, 18, 22 and 26 Hz$^{-1/2}$, while the green cross shows the fitted 95 $\%$ efficiency point. The orange trace shows the linear fit, while the two surrounding traces show the 1-$\sigma$ envelope, from which the 1-$\sigma$ error for the 95$\%$ point is obtained.}
\label{fig:Depth1}
\end{figure}

Figure \ref{fig:Depth2} shows the $95\%$ sensitivity depth points found for the 10 0.1 Hz frequency bands, the mean calculated from these 10 points and two sets of green and red lines, which mark one and two standard deviations respectively. The $95\%$ sensitivity depth point for an extra set of 10 frequency bands has been calculated in order to validate the results. All the new points fall inside the range given by the two standard deviation range. Figure \ref{fig:Depth1} shows an example of the linear fit found at a particular frequency band.

\begin{figure}[]
\includegraphics[width=1.0\columnwidth]{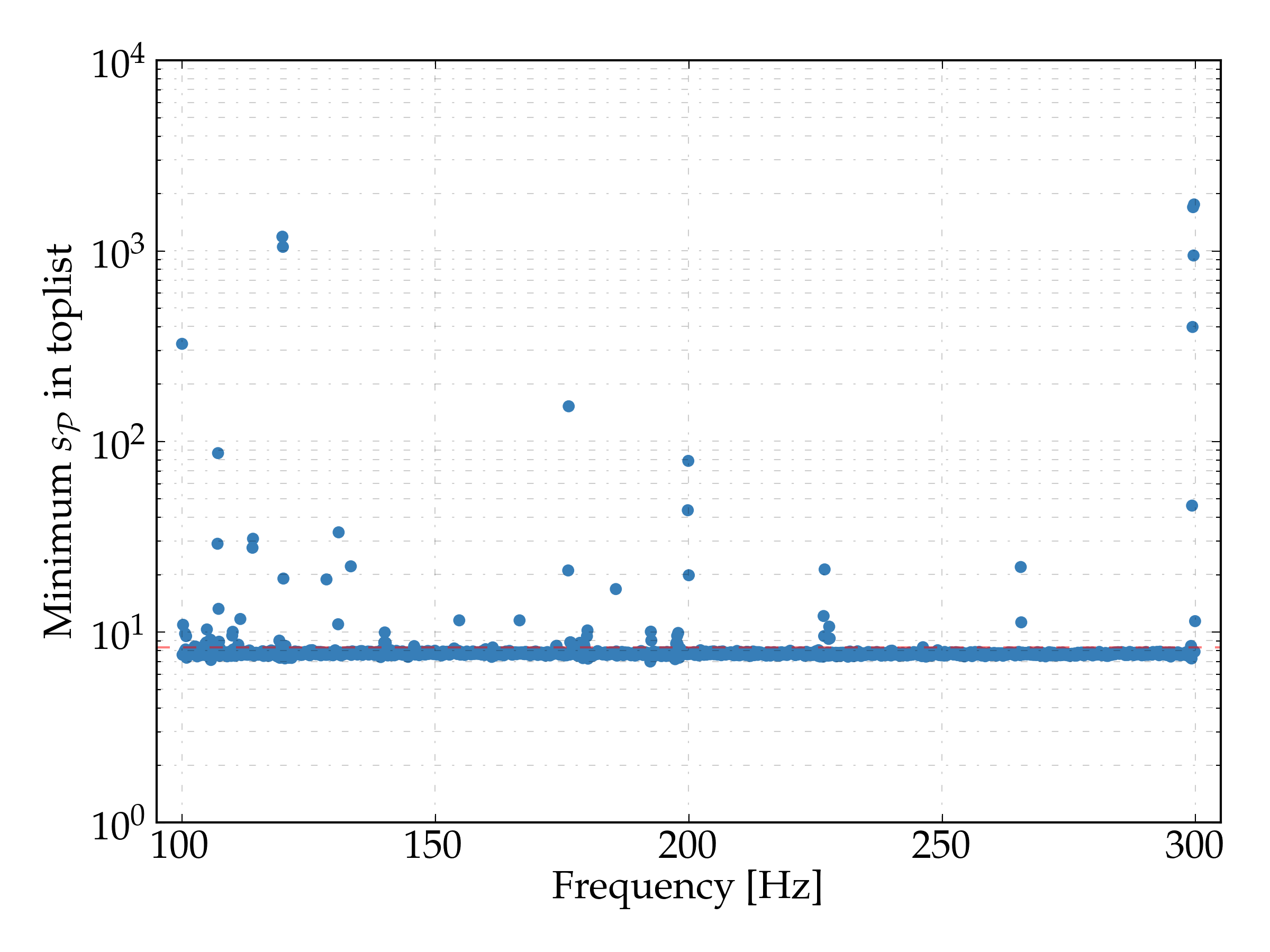}
\caption{Minimum value of the detection statistic obtained in the final toplist for each 0.1 Hz frequency band. The dashed red line at 8.3 marks the threshold.}
\label{fig:Bands}
\end{figure}

\section{Valid upper limits frequency bands}

As mentioned in the main paper, the $95\%$ sensitivity depth value is only valid at 0.1 Hz frequency bands which show a Gaussian behaviour. As the non-Gaussianity of a band increases, the values of the detection statistic also increase, making the detection of a signal producing the same detection statistic harder than in a regular band, since the final toplist will not include it. Bands with non-Gaussian behaviour are usually produced by lines and combs as discussed in the main text or by short-duration glitches which elevate the noise floor of the detector over a wide frequency band.  

A method to localize such frequency bands is to plot the minimum value of the detection statistic $s_{\mathcal{P}}$ for all frequency bands, as figure \ref{fig:Bands} shows. We set an empirical threshold of 8.3, marked with a red line in the plot. 
75 out of the 2000 bands have a value higher than 8.3, and we claim that for these bands the sensitivity depth might be smaller than the value found for Gaussian bands, although we do not present a quantitative study of this effect.

These are the 0.1 Hz non-Gaussian frequency bands: [100.0, 100.2, 100.6, 100.8, 102.5, 102.7, 104.6, 104.8, 104.9, 105.6, 107.0, 107.1, 107.2, 107.3, 109.9, 110.0, 111.1, 111.5, 113.9, 114.0, 119.2, 119.5, 119.8, 119.9, 120.0, 120.4, 128.5, 130.8, 130.9, 133.3, 139.9, 140.0, 140.2, 145.8, 154.7, 161.3, 166.6, 173.8, 173.9, 176.2, 176.3, 176.6, 176.7, 178.5, 179.3, 179.4, 179.9, 180.0, 185.6, 192.5, 192.6, 197.5, 197.6, 197.7, 197.8, 197.9, 199.8, 199.9, 200.0, 226.6, 226.7, 226.8, 227.6, 227.7, 227.8, 246.2, 265.5, 265.6, 299.1, 299.3, 299.4, 299.5, 299.6, 299.7, 299.9].

\section{Table of outliers}

See table \ref{tab:SkyHoughOutliers}.

\begin{table*}[]
\begin{center}
\begin{tabular}{ r r r r r r r r}
\hline
\multicolumn{1}{c}{Frequency} & \multicolumn{1}{c}{$\alpha$} & \multicolumn{1}{c}{$\delta$} & \multicolumn{1}{c}{$a_p$} & \multicolumn{1}{c}{P} & \multicolumn{1}{c}{$t_{asc}$} & $2\mathcal{F}_{sum}$ & \multicolumn{1}{c}{Description} \\
\multicolumn{1}{c}{[Hz]} & \multicolumn{1}{c}{[rad]}    & \multicolumn{1}{c}{[rad]}     & \multicolumn{1}{c}{[s]} & \multicolumn{1}{c}{[s]} & \multicolumn{1}{c}{[GPS]} & & \\
\hline \hline
100.22026 & 0.013 & -0.608 & 13.90 & 3913534 & 1176008851 & 2027.90 & Short disturbance in H1 \\
100.21259 & -1.205 & -0.585 & 14.26 & 3316110 & 1175772093 & 1992.96 & Short disturbance in H1 \\
100.22268 & 1.639 & -0.857 & 13.42 & 1833880 & 1176635113 & 1963.22 & Short disturbance in H1 \\
103.09979 & 2.718 & 0.051 & 24.74 & 3824964 & 1175503115 & 2003.03 & Short disturbance in L1 \\
145.79107 & -2.818 & 0.498 & 21.59 & 3429767 & 1174617540 & 1987.54 & Hardware injection \\
145.86610 & -2.410 & 0.797 & 11.98 & 3142246 & 1174797410 & 1978.52 & Hardware injection \\
145.90148 & 2.285 & 1.300 & 9.71 & 3671138 & 1176001528 & 1982.27 & Hardware injection \\
145.92549 & 2.487 & 1.335 & 9.62 & 3349536 & 1175854186 & 1981.71 & Hardware injection \\
166.66391 & 1.402 & -1.459 & 10.32 & 2295263 & 1176280470 & 2059.17 & Short disturbance in H1 \\
166.64618 & -1.237 & -1.102 & 12.48 & 2228943 & 1175181758 & 1974.85 & Short disturbance in H1 \\
173.89725 & -0.263 & -1.045 & 21.01 & 3229318 & 1177345010 & 3254.80 & Short disturbance in H1 \\
173.87984 & -0.354 & -0.766 & 34.48 & 1764911 & 1175470277 & 3436.08 & Short disturbance in H1 \\
173.89072 & -0.458 & -0.944 & 28.90 & 2411275 & 1176754746 & 3383.46 & Short disturbance in H1 \\
173.90550 & -0.302 & -0.879 & 13.32 & 2246880 & 1175817043 & 3411.56 & Short disturbance in H1 \\
173.91856 & -0.441 & -0.632 & 27.44 & 1671568 & 1176760037 & 3339.30 & Short disturbance in H1 \\
173.90720 & 0.084 & -0.671 & 18.69 & 1723682 & 1176402358 & 3211.92 & Short disturbance in H1 \\
176.68644 & 2.112 & -0.876 & 33.65 & 1517400 & 1176111051 & 2561.38 & Short disturbance in H1 \\
176.66651 & 2.186 & -0.923 & 32.46 & 1633407 & 1176627316 & 3021.95 & Short disturbance in H1 \\
176.68301 & 3.056 & -0.616 & 32.97 & 1501423 & 1176061869 & 2890.82 & Short disturbance in H1 \\
176.69951 & 2.568 & -0.906 & 23.70 & 1957761 & 1176132478 & 3096.21 & Short disturbance in H1 \\
176.71434 & 2.319 & -0.835 & 33.67 & 1408988 & 1175655939 & 2943.29 & Short disturbance in H1 \\
176.70635 & 2.273 & -0.852 & 35.86 & 2278473 & 1176412120 & 3061.56 & Short disturbance in H1 \\
182.02284 & 2.498 & -0.156 & 18.15 & 3410134 & 1175957444 & 1962.52 & Short disturbance in L1 \\
182.02969 & -1.035 & -0.182 & 31.57 & 2480680 & 1176772000 & 2071.28 & Short disturbance in L1 \\
182.03574 & -0.933 & -0.128 & 18.23 & 2109858 & 1177216961 & 2105.25 & Short disturbance in L1 \\
182.62066 & 1.298 & 1.317 & 28.20 & 2041602 & 1176047775 & 1952.68 & Short disturbance in L1 \\
185.68991 & -0.789 & 0.054 & 12.48 & 3641071 & 1174370500 & 3376.22 & Short disturbance in L1 \\
185.68562 & -1.285 & -0.012 & 18.75 & 2824333 & 1176879812 & 3125.18 & Short disturbance in L1 \\
185.68693 & -1.074 & 0.118 & 19.24 & 2941083 & 1177477791 & 3248.15 & Short disturbance in L1 \\
185.70359 & -0.592 & -0.026 & 29.14 & 3721385 & 1176232839 & 3348.67 & Short disturbance in L1 \\
185.70019 & -0.833 & -0.110 & 21.53 & 3050708 & 1177293756 & 3385.85 & Short disturbance in L1 \\
185.70054 & -0.217 & 0.100 & 10.33 & 2492984 & 1177394628 & 3375.33 & Short disturbance in L1 \\
190.54188 & -0.187 & -0.530 & 17.26 & 3815778 & 1176611444 & 2452.27 & Hardware injection \\
190.58117 & -0.363 & -0.446 & 15.56 & 3738116 & 1176003880 & 2347.43 & Hardware injection \\
190.64508 & -1.080 & -0.185 & 9.78 & 2901531 & 1177606154 & 1999.52 & Hardware injection \\
190.69765 & 0.044 & 0.051 & 12.33 & 3354700 & 1177521960 & 2261.82 & Hardware injection \\
190.69851 & 0.167 & 0.039 & 13.66 & 3621201 & 1174548675 & 2383.19 & Hardware injection \\
\hline
\end{tabular}
\caption{Outliers found by this search. All of them can be ascribed either to a hardware injection or to a noise disturbance. The parameters correspond to the center of the cluster returned by the follow-up stage. The $2\mathcal{F}_{sum}$ column shows the summed semi-coherent $\mathcal{F}$-statistic over segments of the top candidate obtained at the first stage of the follow-up. The reference time for these parameters is 1164562334 GPS.}
\label{tab:SkyHoughOutliers}
\end{center}
\end{table*}